\documentclass[%
 reprint,%
 amssymb, amsmath,%
 aip,cha,%
]{revtex4-1}

\usepackage{bm}%
\usepackage[colorlinks=true,linkcolor=blue]{hyperref}%
\expandafter\ifx\csname package@font\endcsname\relax\else
 \expandafter\expandafter
 \expandafter\usepackage
 \expandafter\expandafter
 \expandafter{\csname package@font\endcsname}%
\fi
\usepackage[utf8]{inputenc}
\usepackage{graphicx}
\usepackage{dcolumn}
\usepackage{bm}
\usepackage{hyperref}
\usepackage{physics}
\usepackage{amsmath}
\usepackage[mathlines]{lineno}
\usepackage{xcolor}
\usepackage{float} 

\begin{document}

\title{Equation of state of atomic solid hydrogen by stochastic many-body wave function methods} 

\author{Sam Azadi}

\affiliation{Department of Physics, King's College London, Strand, WC2R 2LS London, United Kingdom}
\email{sam.azadi@kcl.ac.uk }

\author{ George H. Booth}
\affiliation{Department of Physics, King's College London, Strand, WC2R 2LS London, United Kingdom}

\author{Thomas D. K\"{u}hne}
\affiliation{Department of Chemistry, Paderborn Center for Parallel Computing, Paderborn University, 33098 Paderborn, Germany}
\date{\today}

\begin{abstract}
 We report a numerical study of the equation of state of crystalline body-centered-cubic (BCC) hydrogen, tackled with a variety of complementary many-body wave function methods. These include continuum stochastic techniques of fixed-node diffusion and variational quantum Monte Carlo, and the Hilbert space stochastic method of full configuration-interaction quantum Monte Carlo. In addition, periodic coupled-cluster methods were also employed. Each of these methods is underpinned with different strengths and approximations, but their combination in order to perform reliable extrapolation to complete basis set and supercell size limits gives confidence in the final results. The methods were found to be in good agreement for equilibrium cell volumes for the system in the BCC phase, with a lattice parameter of 3.307 Bohr.
 \end{abstract}

\maketitle

\section{Introduction}
A stochastic description of quantum mechanics has significant advantages in the understanding of quantum systems, especially when a large number of degrees of freedom are involved. The main advantage of this approach relies on the exploitation of well-established mathematical bounds derived from probability theory and stochastic processes to control the convergence of these properties. In this picture, quantum particles move along stochastic trajectories, and expectation values can be formulated as ensemble averages over the space of these trajectories. In practice, we exploit the similarity between the Schr\"{o}dinger equation in imaginary time, which is a linear, parabolic partial differential equation, and the diffusion equation. The efficiency of Monte Carlo approaches relies on the use of random numbers to sample the $3N$-dimensional phase-space space of configurations, where $N$ is the number of variables \cite{Landau, Kalos, Ceperley95}. 

Quantum Monte Carlo (QMC) approaches to large, {\em ab initio} systems with realistic many-body Hamiltonians have provided some of the most accurate and reliable descriptions of both Fermionic and Bosonic quantum matter \cite{Landau, Becca, Matthew}. The most established QMC techniques include variational Monte Carlo (VMC) \cite{VMC, Umrigar007}, and diffusion Monte Carlo (DMC) \cite{Anderson, Ceperley80}, have been successfully applied to a variety of realistic quantum materials, including chemical systems \cite{Reynold, Grossman, Ozone, Benzene}, and solids \cite{Kolorenc, Dubecky, SignSWF, Disulfide}. The key advantages of these QMC methods is their efficient large-scale parallelization, low scaling with system size, and predictive power. At present, DMC is the most commonly used QMC technique for high-precision wave function-based predictions of material properties \cite{Luke, Hydrogen, Wagner, Antonietti}. DMC works with an ensemble of (almost) independent 'walkers' or 'configurations', whose stochastically realised dynamics ensures that they explore the Fermionic configuration space of the system. Beginning with a trial wave function, samples of the ground state wave function of the system are projected onto by applying the imaginary time operator $\exp(-\tau \it{H})$. The dominant open problem preventing the exact numerical calculation of many-electron systems by DMC is the Fermionic sign problem, which arises from the antisymmetric nature of many-body wave function with respect to electron exchange. A widespread practical solution for this problem is the so-called fixed-node (FN) approximation \cite{AndersonFN}. In this, a many-electron nodal surface is defined as the one coinciding with that given by a trial many-electron wave function. This trial state is a function of $3N$ variables and the trial nodal surface is the $(3N-1)$-dimensional hypersurface on which the function is zero, and across which the wave function amplitude changes sign. The approximation ensures that the hypersurface of the sampled wave function in DMC coincides with that of the trial wave function. This FN algorithm gives the lowest-energy many-electron state with the given nodal surface \cite{Ceperley91}, which ensures that it is a variational approximation. In principle, this approximation would be exact if the applied nodal boundaries coexisted with the exact nodal surface of the many-Fermionic wave function. In practice, however, the errors in FN DMC energies are usually about $5\%$ of the correlation energy for commonly used trial states. 

An alternative approach, to tackle the Fermionic sign problem, is to allow the exact nodal surface to be an emergent property of the underlying algorithm. Such a simulation would therefore not require initial information of a trial nodal surface. It was demonstrated that such property can be obtained in full configuration-interaction quantum Monte Carlo (FCIQMC) \cite{GB009, Booth10, Cleland, Booth11, Booth13}. This technique can be applied to Fermionic systems after projection into a discrete basis set familiar to conventional quantum chemistry approaches. It was demonstrated that this method can converge to capture the complete correlation energy, or full configuration-interaction (FCI) solution, for the given basis set. There are a number of similarities and differences between the DMC and FCIQMC approaches to stochastic realisation of quantum systems. Both techniques run a long-time integration of the imaginary time Schr\"{o}dinger equation. However, while DMC works in continuous real space, FCIQMC spans the Hilbert space of Slater determinants. In DMC walkers follow the diffusion equation, while in FCIQMC the propagation step is based on a fictitious population dynamics of creation and death processes. In DMC and FCIQMC, the wave function is rendered by walkers representing a specific configuration in their respective configurations, which enables the methods to stochastically sample the wave function without storing the exponential number of amplitudes in the space. The key step of the FCIQMC algorithm, which allows the nodal structure of the wave function to emerge, is walker annihilation. Since each walker has a defined sign (or phase for complex-valued wave functions), if two configurations with opposite signs simultaneously occupy the same determinant, both walkers are deleted from the simulation. The walker annihilation mechanism has also been explored in DMC and Green's function Monte Carlo \cite{Arnow, Ceperley84}.

Solving realistic many-body Hamiltonians is the main challenge in condensed matter physics and quantum chemistry. Traditional quantum chemical methods, including coupled-cluster (CC) theory \cite{Purvis, Bartlett81, Bartlett07, Stanton, Cizek, Jeziorski} and configuration-interaction \citep{Sherrill1999}, have been developed for solving the Schr\"{o}dinger equation, primarily for chemical systems described by a given one-particle basis set. These approaches truncate the wave function to a specific ansatz, which can be relaxed to define a systematic hierarchy of approximations to exactness. Due to the computational complexity of these quantum chemical methods, applying this systematic hierarchy of methods for extended systems and solids is in its early stages of research\cite{Muller, Marsman, Gruber, Pulkin20, Gao20, McClain, Booth16}. The application of FCIQMC to realistic solids also demonstrated a promising route for providing reference ground state many-electron energies to benchmark quantum-chemical techniques, including the CC ansatz. On the other hand, many alternate approaches for dealing with the high dimensionality of real extended systems have been developed, including local truncation, single-particle Green's function methods, novel Monte Carlo algorithms, and embedding techniques, all of which also benefit from comparison to higher accuracy approaches rather than experiment \cite{SignSWF, Setten, H2SWF, Lejaeghere, Motta, Williams}. 

In this work, we compute the equation of state (EOS) of atomic solid hydrogen in the body centered cubic (BCC) lattice using CC, FCIQMC and DMC techniques. Studying the BCC phase of solid hydrogen is critical to understand the origin of metallic magnetism for this system. Many theoretical and numerical investigations have concentrated on this atomic phase of solid hydrogen\cite{Ceperley87, Ashcroft89, Mao89, Barbee, Wang90, Natoli, McMahon, PRL14, HydrogenBCC, KeLiao, PRB19}. Despite the fact that at low densities the BCC atomic phase is not the most stable structure of solid hydrogen\cite{Natoli}, investigations on this simple but realistic system can provide qualitative insights into metal-insulator transitions, and also supply a reference for commonly used mean-field approximations. This is why we consider the ground state energy of the paramagnetic BCC phase of solid hydrogen within the density regime $1.3<r_S<2.4$, where the paramagnetic phase is more stable than antiferromagnetic and ferromagnetic states. We compare our FCIQMC results with CC singles-doubles (CCSD) calculations, which are obtained using the same basis set and system size. We then extend our investigation to include DMC to allow for larger system sizes and without a substantial basis set dependence. Since DMC results are not constrained by basis size, the DMC and FCIQMC values can not be directly compared due to their different model assumptions. However, by careful extrapolation of their intrinsic model assumptions, these complementary descriptions can be combined to provide accurate estimates of the correlation energy in the thermodynamic and complete basis set (CBS) limits.

\section{\label{CD}Computational methods}
We briefly describe the FCIQMC and DMC approaches. Both methods have been presented in more detail in previous works \cite{Matthew, Ceperley, GB009}, but less often alongside each other.

\subsection{\label{fci}FCIQMC Method}

In FCIQMC, we first choose a basis of $2M$ one-particle spin-orbitals $\phi_n$, from which the space of all possible $N$-electron determinants can be constructed via 
\begin{eqnarray}\nonumber
D_{\bf i} &=& D_{i_1, ..., i_N} = \frac{1}{\sqrt{N!}} | \phi_{i_1} \cdots \phi_{i_n} | \\
 &=& \frac{1}{\sqrt{N!}} 
\begin{vmatrix}
\phi_{i_1}(r_1) & \cdots & \phi_{i_1}(r_N) \\
\phi_{i_2}(r_1) & \cdots & \phi_{i_2}(r_N) \\
\cdots & \cdots & \cdots \\
\phi_{i_N}(r_1) & \cdots & \phi_{i_N}(r_N) \\
\end{vmatrix}.
\end{eqnarray}
With this set of determinants as an $N$-body basis, the wave function can be expanded as
\begin{equation}
| \Psi \rangle  = \sum c_{\bf i} | D_{\bf i} \rangle ,
\end{equation}
with the optimal coefficients $c_{\bf i}$ being the ones that variationally minimise
\begin{equation}
E(c_{\bf i}) = \frac{\langle \Psi|\widehat{H}|\Psi \rangle}{\langle \Psi|\Psi \rangle} .
\end{equation}
These coefficients are found from the solution of the matrix eigenvalue problem 
\begin{equation}
\sum_j H_{{\bf ij}} c_{\bf j} = E_0^{FCI} c_{\bf i} ,
\label{FCI-eigen}
\end{equation}
where $ H_{{\bf ij}} = \langle D_{\bf i}|H|D_{\bf j} \rangle$. This FCI approach captures all possible correlation energy within the basis of orbitals, and is therefore in principle systematically improvable as the basis is enlarged, albeit at exponential cost. Due to the basis incompleteness error, the calculated total energies are therefore often higher than FN-DMC energies, but cancellation of errors is in general more reliable in appropriately optimized quantum chemical basis sets. However, the total number of determinants increases exponentially with system and basis size, hence the appeal of a stochastic approach which can deal with such high-dimensional spaces. 

The FCIQMC algorithm converges to the FCI (lowest energy) eigenvector of the FCI matrix eigenproblem of Eq.~\ref{FCI-eigen}, via the solution of the imaginary-time Schr\"{o}dinger equation. The master equation governing the stochastic walker dynamics is given by
\begin{eqnarray}\nonumber
\frac{dc_{\bf i}(t)}{dt} &=& -\sum_{\bf j} [H_{\bf{ij}} -(E_{HF} +S) \delta_{\bf{ij}}] c_{\bf j}(t) \\
&=& -\sum_j (K_{\bf{ij}} - S\delta_{\bf{ij}}) c_{\bf j}(t) .
\end{eqnarray}
The amplitude of each determinant is then coarse-grained with a variable resolution, with the vast majority of amplitudes then represented at any single FCIQMC iteration by a zero amplitude. By representing the wave function at any single iteration just by the walkers (necessarily with non-zero weights) rather than the amplitudes, there is a significant compression of the wave function storage requirement to describe this snapshot of the wave function, which is formally decoupled from the size of the underlying Hilbert space. The success of the algorithm therefore relies on the ability to perform this compression, whilst maintaining a faithful realisation of the state, ultimately exploiting its inherent sparsity afforded by the choice of representation. The (signed) population of walkers then sample the configuration space $q_{\bf i}$, updating each iteration depending on a choice of time step $\Delta t$, through a series of `spawning' steps onto connected configurations and `death' steps, which generally reduce the local population of walkers.
These spawning and death steps stochastically update the amplitude on a determinant via
\begin{equation}
q_{\bf i}^{(n+1)} = q_{\bf i}^{(n)} - \Delta t \sum_{\bf j} (K_{{\bf ij}} -S\delta_{{\bf i,j}}) q_{\bf j}^{(n)} ,
\end{equation}
where $q_{\bf i}$ is the random variable denoting the instantaneous walker population on configuration $|D_{\bf i} \rangle$.
These walker dynamics are not dissimilar to a first-order approximation to those of DMC, where the propagator is $\tilde{G} = I - (H - SI) \Delta t$ instead of $G = \exp(-(H-SI)\Delta t)$, but without importance sampling and the FN approximation.
As long as $\Delta t \leq 2/(E_{max} - S)$, FCIQMC will formally yield the exact ground sate without any time-step error.  The energy can be extracted as 
\begin{eqnarray}
E(t) &=& \frac{\langle \exp(-t\hat{H}) D_0|\hat{H}|D_0\rangle}{\langle \exp(-t\hat{H})D_0|D_0\rangle} \\
      &=& E_{HF} + \sum_{{\bf j} \neq 0} \frac{c_{\bf j}(t) \langle D_{\bf j}|\hat{H}|D_0\rangle }{c_0(t)} \\
      &=& E_{HF} + \frac{\langle \sum_{{\bf j} \neq 0} q_{\bf j}(t) H_{{\bf j}0} \rangle}{\langle q_0(t) \rangle} ,
\end{eqnarray}
where $|D_0 \rangle$ represents a trial state \cite{Petruzielo, Blunt15}.

In keeping with all general Fermionic QMC techniques, there is also a sign problem that afflicts FCIQMC \cite{JSpencer}. In order to ensure that the annihilation is sufficient in order to allow the exact nodal structure of the FCI wave function to emerge, the number of walkers must be large enough, and while this number is generally much smaller than the dimensionality of the entire Hilbert space, it still grows exponentially with system size. To reduce the number of walkers required for high accuracy, the FCIQMC algorithm can be improved with the `initiator' approximation \cite{Cleland}. Initiator FCIQMC is a systematically improvable approximation \cite{Booth10, Booth11}, whereby the determinants are divided into two classes, labelled `initiator' and `non-initiator'. Initiator determinants are allowed to create new walkers on unoccupied determinants. However non-initiator determinants are only allowed to create new walkers on already occupied determinants. The label of initiator or non-initiator is chosen depending on the current population of walkers on any determinant, with the threshold $n_{\rm add}$ used to determine this label. This dynamic adaptation helps the walker population to stabilize a sign structure at far lower walker numbers, at the cost of introducing a small systematic error. This error can be systematically reduced as the number of walkers is increased, which enlarges the set of initiators. Other adaptations in recent years have also improved the accuracy and scope of the method, including the computation of excited states\cite{Blunt2017, Blunt2015}, unbiased molecular properties \cite{Overy14, Thomas15, Samanta18} and its use as a solver within an active space framework\cite{Thomas2015, Anderson2020}. 

\subsection{\label{dmc}Diffusion Monte Carlo}
The DMC method is a stochastic technique for many-electron systems with a much longer history than FCIQMC. Any solution of the time-dependent Schr\"{o}dinger equation can be expanded in the eigenfunctions of the Hamiltonian ${\Phi_n({\bf R})}$, i.e.
\begin{equation}
\Psi ({\bf R}, \tau) = \Sigma_n c_n \Phi_n ({\bf R}) \exp(-E_n\tau), 
\end{equation}
where $E_n$ is the eigenvalue corresponding to $\Phi_n$ and $\tau = it$. One can find that $\Psi ({\bf R}, \tau \rightarrow \infty) \simeq c_0 \Phi_0 ({\bf R}) \exp(-E_0 \tau)$, which is proportional to the ground state wave function. In principle, the Schr\"{o}dinger equation can be solved by propagating an arbitrary wave function in imaginary time for long enough. This propagation can be seen in the differential form of the imaginary-time Schr\"{o}dinger equation
\begin{eqnarray}\nonumber
&&-\frac{1}{2} \nabla^2 \Psi ({\bf R}, \tau) + [V({\bf R}) -E_T] \Psi({\bf R}, \tau) \\
&=& -\frac{\partial}{\partial \tau} \Psi({\bf R}, \tau), 
\end{eqnarray}
where $\nabla^2 = \Sigma_i \nabla_i^2$ acts over all coordinates within the vector ${\bf R}$, whereas $E_T$ is a constant energy offset. This equation is equivalent to a diffusion equation, in which $\Psi ({\bf R})$ represents the density of particles at point ${\bf R}$. The particles diffuse with a diffusion coefficient $D = 1/2$, and are absorbed with rate $V({\bf R}) - E_T$. Assuming $\Psi ({\bf R, \tau})$ is a probability density, we distribute an initial set of walkers with probability density given by $\Psi ({\bf R, \tau})$. The walkers then diffuse and can be removed or created accordingly. This can be simulated via a stochastic process, whereby in the limit $\tau \rightarrow \infty$, the walkers would be distributed according to the ground sate wave function. 

For a Fermionic system, a wave function $\Psi({\bf R})$ must have both positive and negative regions to be antisymmetric with respect to particle exchange. Hence, it can not be used as a probability density. This problem can be overcome by using a guiding function $f({\bf R}, \tau) = \Psi_T({\bf R}) \Psi({\bf R}, \tau)$.  Provided $\Psi_T({\bf R})$ and $\Psi({\bf R}, \tau)$ have the same nodal surface, $f({\bf R}, \tau)$ has the same sign over all configuration space, and can be interpreted as a probability distribution function. By multiplying both sides of the imaginary-time Schr\"{o}dinger equation by $\Psi_T$, we obtain
\begin{eqnarray}
-\frac{\partial f}{\partial \tau} = -\frac{1}{2} \nabla^2 f +\nabla \cdot [f{\bf v}] + [E_L - E_T] f,
\label{FP-eq}
\end{eqnarray}
where ${\bf v}({\bf R}) = \nabla \Psi_T ({\bf R})/\Psi_T({\bf R})$ and $E_L({\bf R}) = \hat{H}\Psi_T({\bf R})/\Psi_T({\bf R})$. If we consider $f$ as a probability distribution, this equation is the Fokker-Planck equation describing the diffusion of non-interacting classical particles, with an imposed drift velocity ${\bf v}({\bf R}, \tau)$ and absorption coefficient $[E_L({\bf R}) - E_T]$. We can therefore distribute a set of particles according to an initial distribution $f({\bf R}, 0) = |\Psi_T({\bf R})|^2$, and let them evolve according to the Fokker-Planck equation. In the limit of $\tau \rightarrow \infty$, the walkers will be distributed according to the minimal energy wave function with the same nodal surface as the trial wave function\cite{Umrigar}. The quality of the nodal surface of the trial wave function therefore determines the error of DMC in an uncontrolled way. In practice, the trial wave function is optimised by VMC before being used in DMC. Generally, lower VMC energy imply a better nodal surface \cite{casino, Turbo}. Further details of the implementation of the DMC algorithm are discussed in Ref.~\onlinecite{casino}. 

\subsection{\label{nd}Simulation setup}
FCIQMC calculations were performed using the NECI\cite{NECI} package. The periodic Hartree-Fock (HF) calculations, from which the single particle orbitals were extracted, and CCSD simulations were carried out using PySCF \cite{pyscf1, pyscf2} with norm-conserving pseudopotentials \cite{gthPP}. Gaussian basis sets of SZV, DZVP, and TZVP quality were used for all basis set-based calculations\cite{Vondele}, with density fitting employed in the computation of two-electron matrix elements. For the FCIQMC simulations, up to $1.5\times10^8$ walkers were used for the most demanding calculations, at lower densities and larger simulation cells \cite{NECI}.

Our VMC and DMC calculations were performed using the CASINO QMC package \cite{casino} and a trial wave function of the Slater-Jastrow form was employed. The one-electron orbitals defining the Slater determinant, were extracted from density functional theory (DFT) calculations using the Quantum Espresso code \cite{QE}, with a DFT plane-wave cutoff of $5,000$ eV. The norm-conserving DFT pseudopotential with the Perdew-Zunger parameterization\cite{lda} of the local density approximation was used for our DFT, VMC and DMC calculations. The Jastrow term $J({\bf R})$, which captures most of the dynamical correlation between electrons is a positive, symmetric, explicit function of interparticle distances and consisted of polynomial one-body electron-nucleus (en), two-body electron-electron (ee), and three-body electron-electron-nucleus (een) terms. Whereas the parameters of the Jastrow were optimized by variance minimization at the VMC level \cite{Umrigar88, Neil05}, the Slater determinant was taken from DFT, but not reoptimized. The main approximation in the DMC results is the FN approximation, which can be improved by including backflow transformation in the trial wave function \cite{BF}. The BCC unit cell, which was used to build a supercell for all calculations, includes two hydrogen atoms located at the corner and centre of the cell that were fixed for all the investigated densities and only the lattice parameter was changed for each $r_S$. To minimize the time-step error, a small time step of $\tau = 0.005~a.u.$ was used in all the DMC simulations.

\section{\label{result}Results and discussion}

\subsection{FCIQMC and CCSD}

\begin{figure}[t]
\centering
\includegraphics[width=0.5\textwidth]{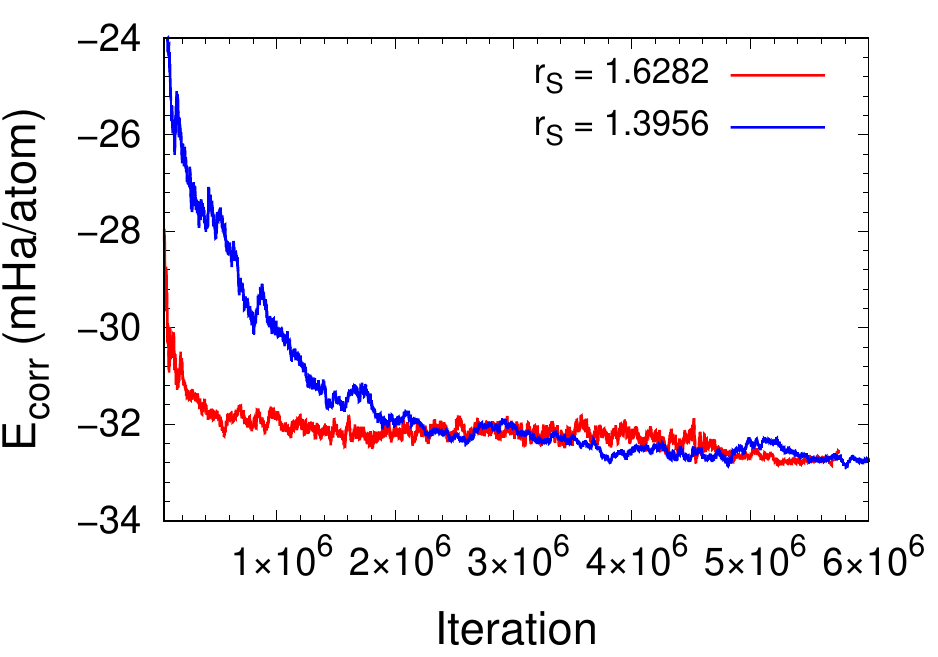}
\caption{\label{Ecorr} (Colour online) Convergence of the FCIQMC correlation energy per atom with respect to the number of iterations. The simulation cell consists of 16 hydrogen atoms using a DZVP basis set and 80 orbitals for two representative densities. Walker number and simulation length were determined in order to surpass mHa/atom precision, which eventually results in 50 million walkers. This maximum walker number was reached for the $r_s=1.3956$ calculation at iteration $5.1\times10^6$, while for the $r_s=1.6282$ calculation, the walker population was grown to $55$ million walkers until iteration $5.0\times10^6$, at which point the initiator error was considered converged, and statistics were accumulated.}
\end{figure}

We first consider the convergence of the FCIQMC correlation energy in a restricted simulation cell. This is given in Fig.~\ref{Ecorr}, where representative convergence of the BCC crystalline hydrogen for two densities is given within a DZVP basis and a simulation cell of 16 hydrogen atoms and 80 orbitals. The total number of walkers in the calculation for each unit cell volume was grown in stages, to check the convergence of the energy estimator. The number of walkers required to achieve convergence with respect to the initiator error varied between the cell sizes, from 40 million walkers at compressed geometries, to 150 million walkers for more expanded geometries, where stronger static correlation effects are expected to be prevalent. Remaining systematic errors are expected to be sub-mHa per atom, resulting in confidence in its ability to benchmark other approaches at these restricted cell and basis sizes.

We also consider the convergence of the energy with respect to the basis set size, considering both the DZVP (with 5 orbitals per atom) and the smaller SZV basis (with only one basis functions per atom). Table~\ref{basis_fci} gives the corresponding FCIQMC energies of the system at different densities within a statistical uncertainty of around 0.1 mH/atom from a blocking analysis, with likely remaining systematic error also sub-mH/atom, giving confidence in the ability of the FCIQMC to be used as a finite-basis benchmark. We compare these results to those of CCSD, obtained using the pyscf simulation package. Table \ref{basis_ccsd} gives the CCSD energies of BCC atomic hydrogen with two atoms per primitive unit cell, which were obtained using SZV, DZVP, and larger TZVP basis sets and a $2\times2\times2$ k-point mesh, at different densities. 
The CCSD total energy at the CBS limit are also given by a three-point extrapolation of the HF energy ($E_{HF}^{CBS}$) via a exponential form, as well as the CCSD correlation energy ($\Delta E_{CCSD}^{CBS}$), via the standard two-point (DZVP and TZVP) inverse-cubic form to the complete basis set (CBS) limit\cite{Feller, Helgaker, Tajti}.  
The values of the CCSD total energy in the CBS limit given by $E_{tot}^{CBS} = E_{HF}^{CBS} + \Delta E_{CCSD}^{CBS} $ are reported in Table~\ref{basis_ccsd} as a function of $r_S$. Comparing the FCIQMC-DZVP (Table~\ref{basis_fci}) with the corresponding CCSD energies (Table~\ref{basis_ccsd}) in the same basis set can quantify the systematic error in the latter. The close agreement of these results shows the ability of CCSD to recover the dominant correlated physics in this system, in particular for small $r_S$ values, and underlines the potential of a composite scheme which will be used later, where the lower-cost CCSD can be used to correct the FCIQMC results to account for the effect of remaining basis set incompleteness (or alternatively, for the FCIQMC to correct for static correlation errors in the CCSD).

\begin{table}
\centering
\begin{tabular}{c| c | c  }
$r_S$ & $E_{tot}^{SZV}$ &$E_{tot}^{DZVP}$  \\
\hline
1.3956 & -0.4875(1) & -0.5135(1)  \\
1.6282 & -0.5038(1) & -0.5271(1)  \\
1.8608 & -0.5046(2) & -0.5203(1)  \\
2.0935 & -0.5011(1) & -0.5121(1)   \\
2.3261 & -0.4998(1) & -0.5016(1)   \\
\end{tabular}
\centering
\caption{\label{basis_fci} FCIQMC total energy in Ha/atom as obtained by SZV, and DZVP basis sets at different densities $r_S$. The simulation cell includes 16 hydrogen atoms. Parentheses denote the stochastic error in the last digit as obtained from a blocking analysis.} 
\end{table}

%
\begin{table}
\centering
\begin{tabular}{c| c | c | c | c}
$r_S$ & $E_{tot}^{SZV}$ &$E_{tot}^{DZVP}$ &$E_{tot}^{TZVP}$ & $E_{tot}^{CBS}$\\
\hline
	1.3956& -0.487386 & -0.513529 & -0.513879 &  -0.513983  \\
	1.6282& -0.502846 & -0.524253 & -0.524793 &  -0.524867\\
	1.8608& -0.502244 & -0.518626 & -0.519386 &  -0.519429 \\
	2.0935& -0.496091 & -0.508172 & -0.509355 &  -0.509403 \\
	2.3261& -0.489972 & -0.498029 & -0.499688 &  -0.499938 \\
\end{tabular}
\centering
\caption{\label{basis_ccsd} CCSD total energies in Ha/atom, obtained using SZV, DZVP, and TZVP basis sets at different densities $r_S$. In the final column, the CCSD total energies extrapolated to CBS limit ($E_{tot}^{CBS}$) is shown. The primitive cell includes two hydrogen atoms, sampled with a $2\times2\times2$ k-point grid.}
\end{table}
%

\begin{figure}
\centering
\includegraphics[width=0.5\textwidth]{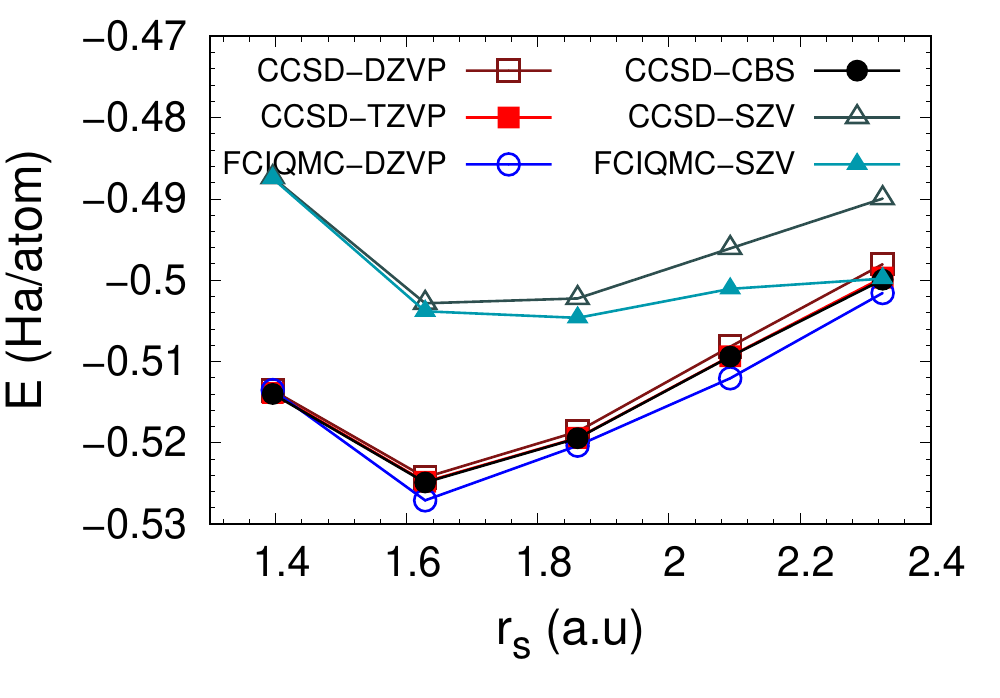} 
\caption{\label{EOS} Total energy per atom obtained by CCSD with SZV, DZVP, TZVP basis sets and at CBS limit, and FCIQMC with SZV, and DZVP basis sets, as the unit cell volume is varied. The same supercell of 16 hydrogen atoms was used for our CCSD and FCIQMC simulations.}
\end{figure}

From these results, we can now obtain the EOS of BCC atomic hydrogen for these methods, both for finite basis sets, and also for extrapolated basis results, within $2\times2\times2$ simulation cells. This is shown in Fig.~\ref{EOS} for the FCIQMC and CCSD methods, respectively. Agreement between these two methods is excellent at higher densities, where the single-reference nature of the CCSD ansatz is expected to perform well. 
Around the equilibrium cell volumes, at $r_S=1.6282$ and $r_S=1.8608$ a.u, the difference between FCIQMC-SZV and CCSD-SZV energies are -0.97, and -2.4 mHa/atom, respectively.  However, for expanded cell volumes, at the $r_S=2.3261$ a.u., the correlation energy captured by FCIQMC increases significantly over the CCSD results for both basis sizes, with the FCIQMC-SZV energy being 9.8 mHa/atom lower than the CCSD-SZV value. This is anticipated, due to the increasing levels of stronger correlation effects present in the system, as evidenced by the increasingly multiconfigurational nature of the FCIQMC calculations.

Using the larger DZVP basis set introduces a major energy gain within the FCIQMC and CCSD calculations. At the equilibrium cell volume, the difference between FCIQMC-DZVP and FCIQMC-SZV is 23.3 mHa/atom. Once again, the agreement between FCIQMC and CCSD at compressed cells is excellent, with the agreement getting worse as the cell expands. However, using the larger basis set, the discrepancy between the CCSD and FCIQMC values for more expanded geometries is much smaller, pointing to an overestimation of the relative importance of strong correlation effects in small basis sets, while larger basis sets are able to more effectively screen these strong correlations. This improved description of the screening available in the larger basis results in a qualitatively different shape to the EOS, reducing the compressibility of the system, and predicting an equilibrium volume of $r_S=1.6282$~a.u., corresponding to a BCC lattice parameter of 3.307 Bohr. 

We were also able to conduct CCSD calculations using the more complete TZVP basis set. Employing this increased basis did not qualitatively change the EOS, with the energies only deviating from those using the DZVP basis by just over $1$ mHa/atom across all considered densities. Furthermore, extrapolating the CCSD energies to CBS limit marginally lowers the total energies compared to the CCSD-TZVP results, with this incompleteness being more notable at larger cell volumes. However, the DZVP basis is still found to be accurate for all cell sizes. 
\begin{table}
\begin{tabular}{l| c| c| c}
Trial state                 & VMC         & Variance  &   DMC \\
\hline
Slater             & -0.47194(5) & 3.01(2)   & -0.50715(1) \\
Slater+2bJ         & -0.50101(3) & 0.6990(9) & -0.507470(6)\\
Slater+2,1bJ       & -0.50110(3) & 0.698(3)  & -0.507471(9)\\
Slater+2,1,3bJ     & -0.50118(3) & 0.719(2)  & -0.507473(6)\\
Slater+BF          & -0.46487(7) & 4.4(1)    & -0.50795(3) \\
Slater+1,2,3bJ+BF  & -0.50429(3) & 0.6244(8) & -0.50851(1)\\
\end{tabular}
\caption{\label{VMC_DMC} VMC and DMC energies of BCC atomic hydrogen at $r_S = 1.3956$~a.u. in a supercell made of 16 hydrogen atoms. The energies for different trial wave functions, given as a Slater determinant without Jastrow term, Slater determinant with two-body Jastrow (Slater+2bJ), with two- and one-body Jastrow (Slater+2,1bJ), and with two-, one, and three-body Jastrow (Slater+2,1,3bJ), Slater determinant with only Backflow (Slater+BF), as well as with Backflow and Jastrow including one-, two-, and three-body terms (Slater+1,2,3bJ+BF).}
\end{table}
\subsection{Diffusion Monte Carlo}
The results of our VMC and DMC calculations at a cell volume of $r_s=1.3956$~a.u. with 16 atoms are shown in Table.~\ref{VMC_DMC}. These show that the dominant error in the DMC values derives from the FN approximation, since the accuracy of the DMC is almost independent of the flexibility afforded by the Jastrow component of the trial wave function, which does not affect the nodal structure. Other systematic errors in DMC, such as time-step errors or non-local pseudopotential errors {\em can} be affected by the quality of the Jastrow component, with these results demonstrating that these errors are small. This suggests that while the importance of the two-body Jastrow is significant for the VMC energies, it may be more efficient to consider a trial wave function for DMC that has not included the 2-body Jastrow terms, which can result is a substantial speedup in the calculation. 

The 1-body Jastrow terms are also found to be negligible in this system, with the 3-body terms also found to be less important to the DMC energy than the inclusion of backflow correlations.
It is found that the backflow reduces the DMC energy by 0.8 mHa/atom, which is more than twice the energy gain originating from the inclusion of the 2-body Jastrow. The importance of backflow in high-density matter has been seen before, where in the homogeneous three-dimensional electron gas with $r_S < 5$ the effects of backflow were found to dominate over those introduced by three-body correlations\cite{Kwon}. Indeed, for this pure hydrogen system, the inclusion of all Jastrow terms and backflow transformation into the wave function results in only a $\sim$1.36 mHa/atom improvement in the DMC total energy compared to just using a single Slater determinant trial state. This energy gain is smaller than what is generally known as chemical accuracy, and so is unlikely to be worthwhile in general for pure hydrogen systems.
\begin{figure}[t]
\centering
\includegraphics[width=0.5\textwidth]{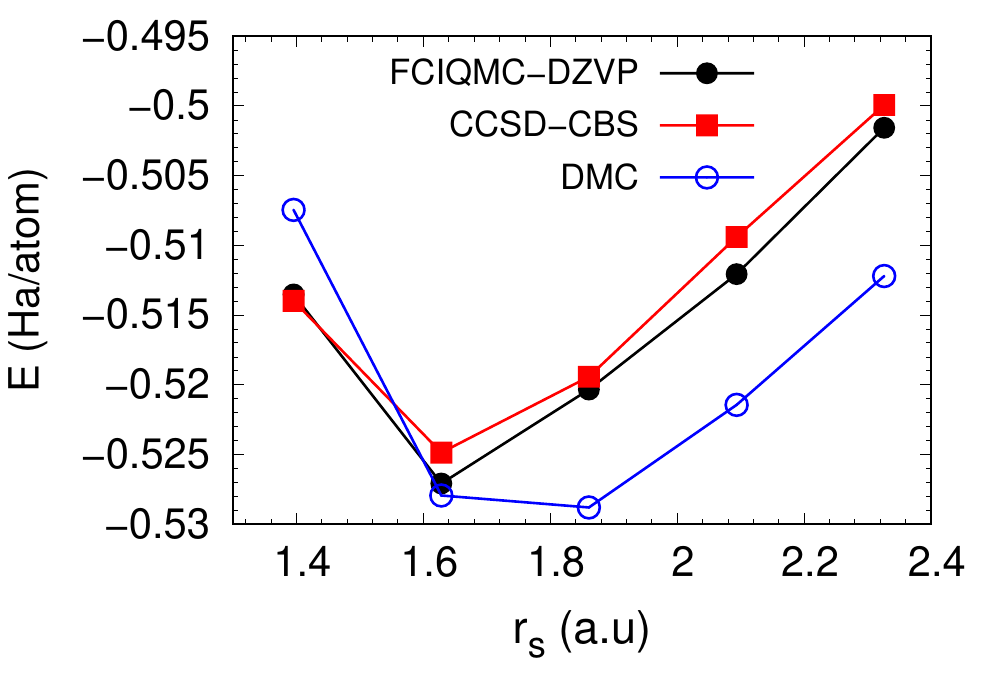}
\caption{\label{N16} (Colour online) FCIQMC-DZVP, CCSD-CBS, and DMC energies as function of $r_S$. The same $2\times2\times2$ supercell size (16 hydrogen atoms) was used for all three methods. The statistical errors within the DMC energies is $\sim 5\times10^{-6}$ Ha/atom. } 
\end{figure}

The comparison of DMC, FCIQMC-DZVP, and CCSD-CBS energies in this reduced 16 atom simulation cell are shown in Fig.~\ref{N16}. We find a qualitative difference between the Fock space approaches of FCIQMC and CCSD compared to the real-space DMC approach. The former methods exhibit a sharper minimum around $r_S=1.6$~a.u., while the DMC EOS has a shallow minimum in the density window of $1.6 < r_S < 1.86$. This discrepancy is not due to basis set or strong correlation effects, but rather the differing rates of convergence of these methods to the thermodynamic limit. It is clear that any comparison and agreement between these methods will have to ensure that the effects of these small supercell sizes are accounted for, which is considered in the following section.

\subsection{Finite size convergence}
Using a finite simulation cell to mimic the properties of a periodic system introduces finite-size (FS) errors, which is one of the main challenges in the application of many-body techniques to extended systems. These FS errors have a number of different origins, including the description of the kinetic energy, the periodic Ewald interaction, and exchange energy \cite{Neil08, FS15, Holzman16, FS19}. The convergence of these different terms to the thermodynamic limit can vary between methods and their representation, as indicated in the previous section. Therefore, careful control to mitigate the impact of finite simulation cells and ensure that all desired quantities are converged with respect to these errors is essential for reliable and comparable results. Here, we employ the standard FS extrapolation technique introduced by Ceperley and co-workers for reducing the FS errors\cite{Ceperley-FS}. Specifically, we employ a $1/N$ extrapolation form of
\begin{equation}
E_{MB, \infty} \sim E_{MB, N} + \alpha (E_{0, \infty} - E_{0, N}) + \beta/N, 
\label{fitting}
\end{equation}
where $\alpha$ and $\beta$ are fitting parameters, and $E_0$ is the energy of system obtained via a single-particle, or mean-field approach. Therein, $E_{MB, N}$ represents the many-body energy of the finite system of $N$ interacting electrons. Using this form, we extrapolate the CCSD-SZV and DMC energies to the infinite system size limit. For the CCSD-SZV calculations, the $E_{0, \infty}$ value is approximated to be the HF energy of the system for that density, obtained using an $8\times8\times8$ k-point mesh. For the DMC extrapolation, the local density approximation was used with a $24\times24\times24$ k-point mesh to obtain $E_{0, \infty}$.

\begin{figure}[h]
\centering
\includegraphics[width=0.5\textwidth]{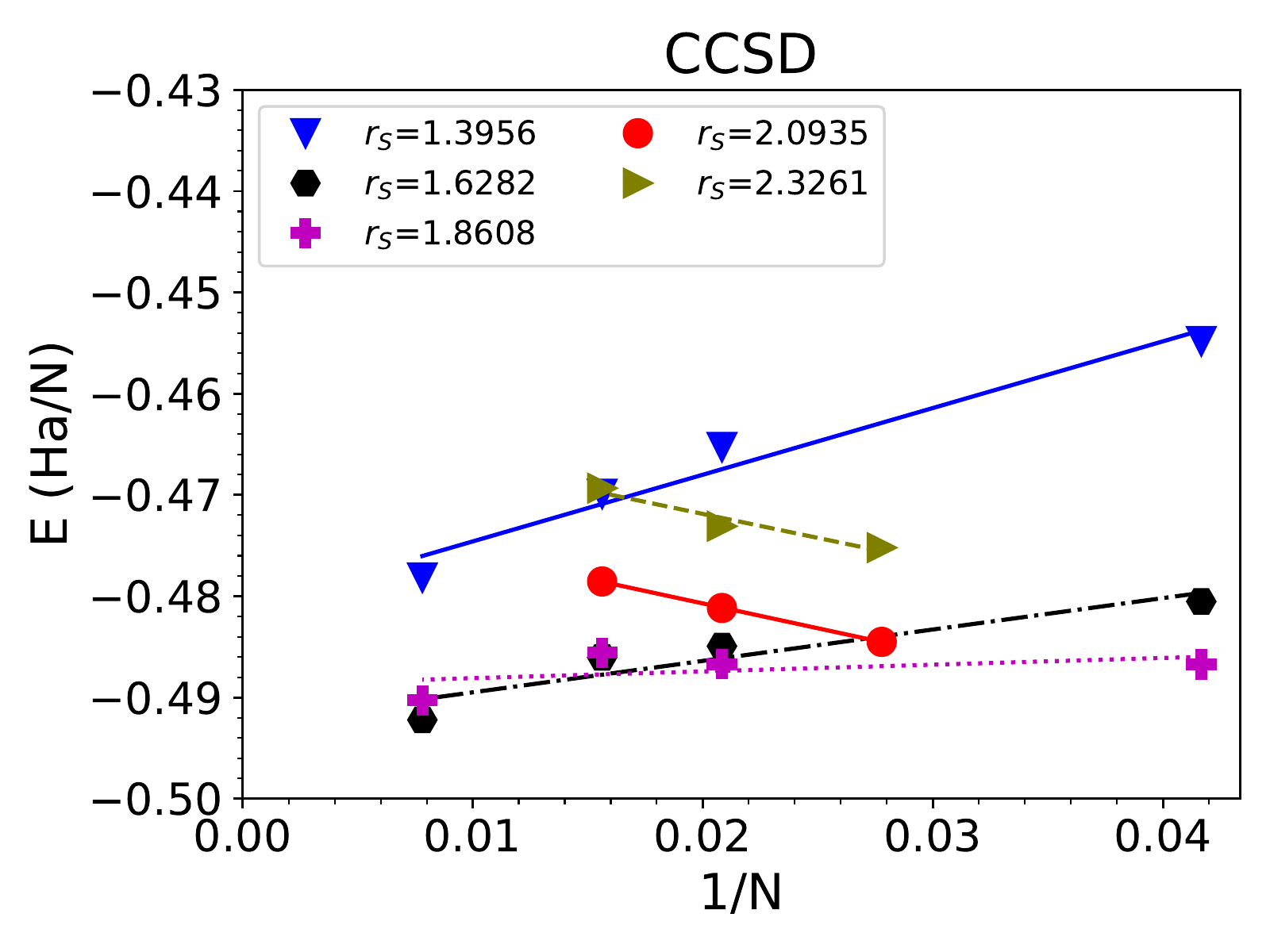}
\includegraphics[width=0.5\textwidth]{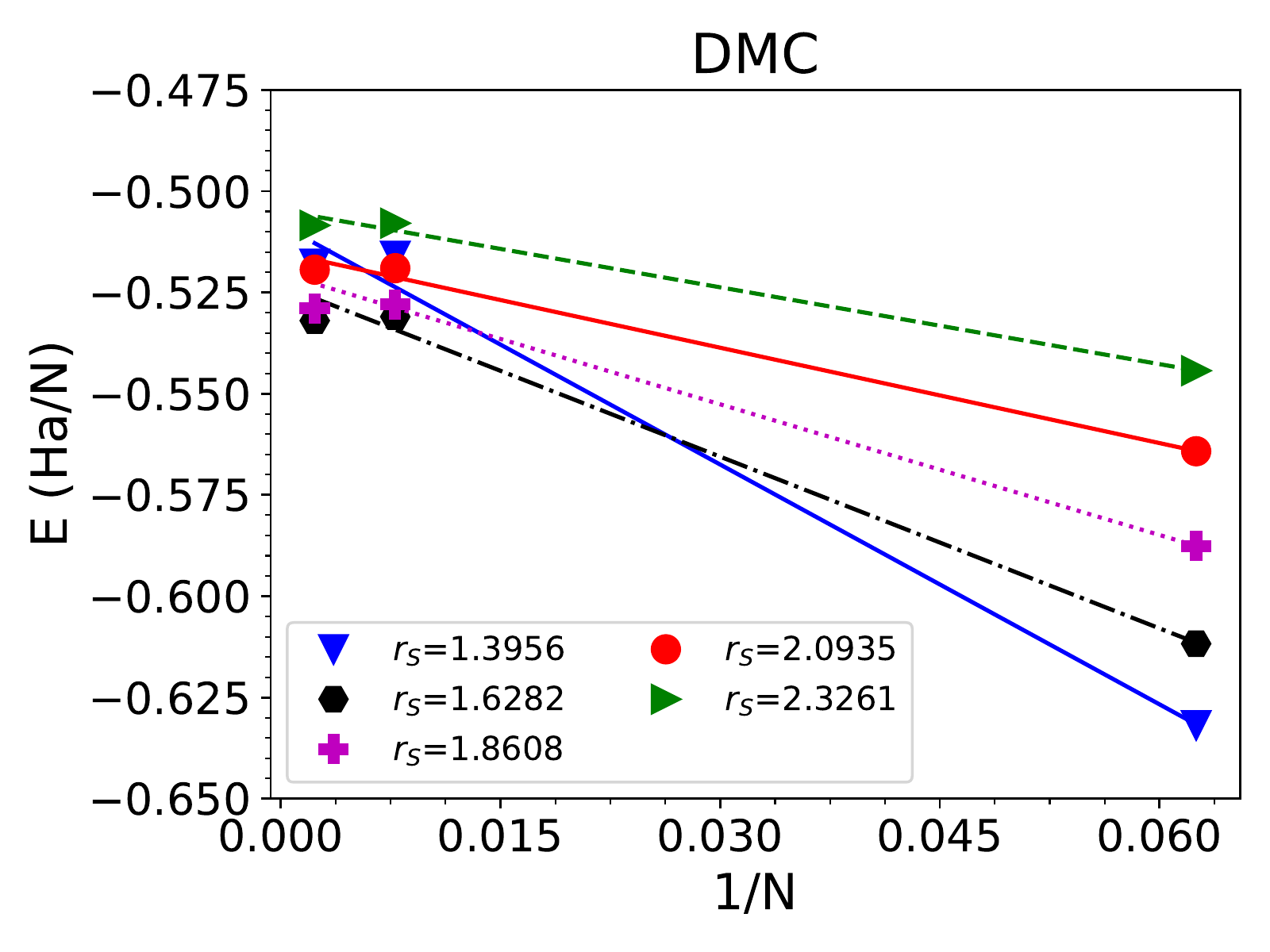}
\caption{\label{FS} Total energies for CCSD within a SZV basis set (top panel) and DMC (bottom panel), as a function of the inverse number of atoms in the supercell, for a range of cell volumes. The DMC energies are obtained employing a large plane wave basis set with an energy cutoff of 5000 eV. A two parameter fit to Eq.~\ref{fitting} is used to extrapolate to the infinite system size limit, given in Table~\ref{E_oo}. } 
\end{figure}

The top panel of Fig.~\ref{FS} shows CCSD-SZV energies, which are calculated at four supercell sizes and different densities, up to a maximum supercell size of $4\times4\times4$. 
The bottom panel shows the extrapolation of the DMC energies at each density (computed with the Slater-Jastrow trial wave function including one- and two-body terms), up to the largest supercell size with 432 atoms. The final CCSD-SZV and DMC energies extrapolated to the infinite system size limit, estimated from Eq.~\ref{fitting}, are given in Table~\ref{E_oo} along with the standard deviation in the fit. 

We find that the finite size correction of the DMC results is always positive, regardless of density. Hence, the DMC energies of finite systems are lower than the DMC energy at the thermodynamic limit. In contrast, the CCSD FS correction lowers the energy of the system for the density range of $r_S < 2.0$, while is positive for lower densities. 
However, the comparison between these approaches is fundamentally limited by the small basis size (SZV) of the CCSD energies used for the extrapolation, which was required in order to reach the large supercell sizes. While this basis is qualitative different to the CBS limit, we aim to use these results as a finite size correction to the more complete basis set results.

\begin{table}
\centering
\begin{tabular}{c| c c | c c}
$r_S$/a.u.  & $E_{CCSD, \infty}$  & $\sigma_{CCSD}$ & $E_{DMC, \infty}$ & $\sigma_{DMC}$  \\
\hline
1.3956 & -0.48117(1)  & 0.135(3) & -0.50827(5)  &  0.145(7) \\
1.6282 & -0.492595(9) & 0.121(3) & -0.52315(2)  & 0.111(4) \\
1.8608 & -0.488757(9) & 0.124(3) & -0.520355(6) & 0.048(3) \\
2.0935 & -0.470905(1) & 0.008(1) & -0.515105(6) & 0.049(2) \\
2.3261 & -0.46240(1)  & 0.165(4) & -0.504769(5) & 0.040(2) \\
\end{tabular}
\caption{\label{E_oo} Extrapolated thermodynamic limit CCSD ($E_{CCSD, \infty}$) and DMC ($E_{DMC, \infty}$) energies and their corresponding standard deviations $\sigma$ of the fit, for each density. The CCSD energies are extrapolated from results using a SZV basis set, whereas the DMC energies are obtained using a Slater+2,1bJ trial wave function. All energies are in Ha/atom, and the number in parentheses provides the variance of the parameter estimate.}
\end{table}

To correct the finite-size extrapolated CCSD results to mitigate for this small basis, we can assume that the basis set error between SZV and DZVP is independent of supercell size, with these errors therefore being additive. This allows us to compute a final basis set correction, which can be applied to the extrapolated CCSD results at the SZV basis level. Furthermore, we can similarly compute the energetic correction between CCSD and FCIQMC in the DZVP basis sets, and also consider this as a correction for the correlated physics beyond the CCSD ansatz in the thermodynamic limit. Leveraging these strengths of these different methods, we are able to compute a final EOS for this composite approach.

\begin{figure}
\centering
\includegraphics[width=0.48\textwidth]{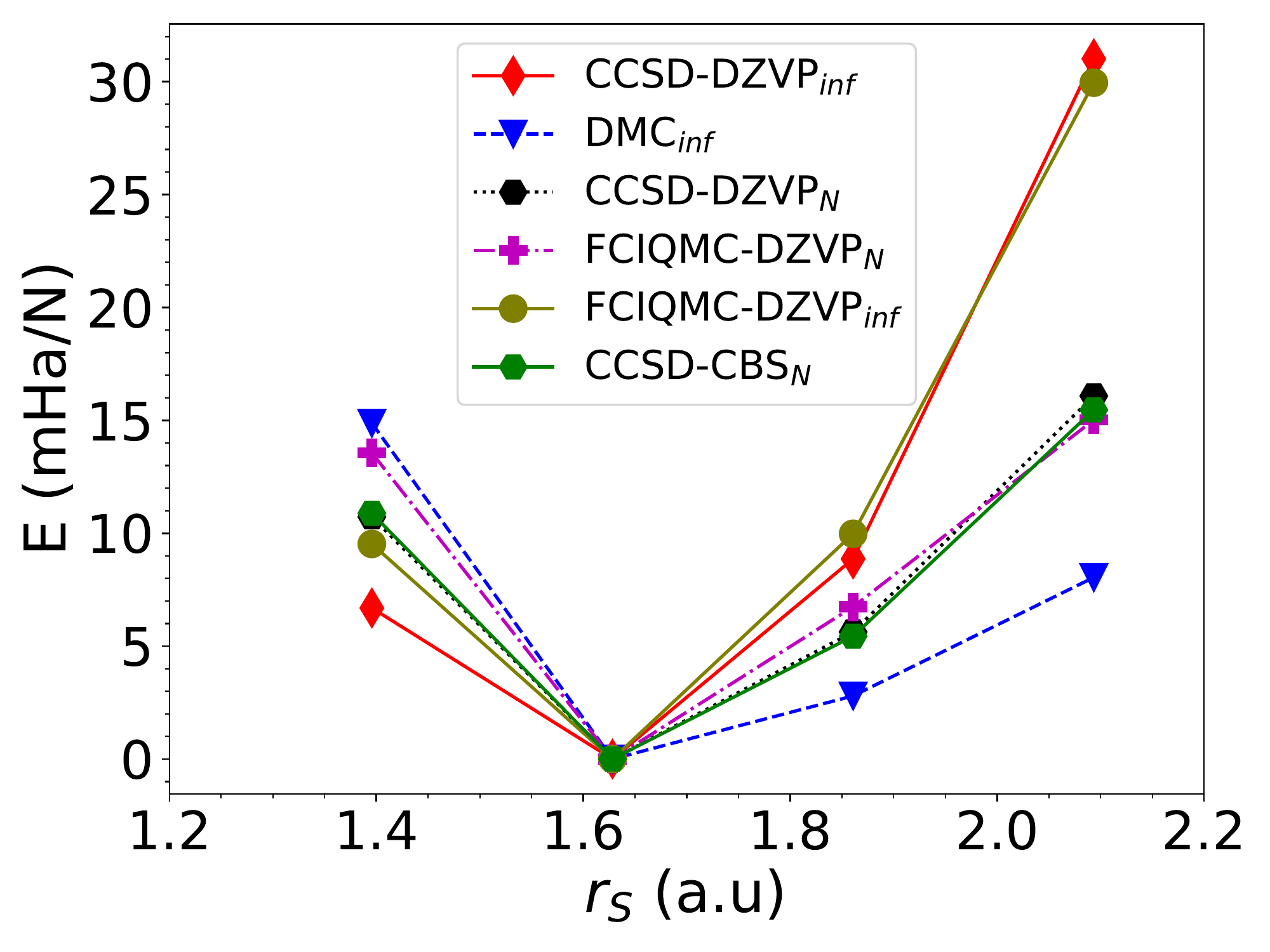}
\caption{\label{E_all} Equation of state for BCC solid atomic hydrogen. FCIQMC-DZVP$_N$ and CCSD-DZVP$_N$ denotes results obtained employing a DZVP basis set and a 16 atom supercell, whereas FCIQMC-DZVP$_{\rm inf}$ and CCSD-DZVP$_{\rm inf}$ are estimated results extrapolated to the infinite system size limit. DMC$_{\rm inf}$ represents DMC energies also extrapolated to the thermodynamic limit.} 
\end{figure}
\subsection{BCC atomic solid hydrogen equation of state}
Figure~\ref{E_all} illustrates the final equation of state for BCC atomic solid hydrogen, as obtained from our FCIQMC, CCSD, and DMC calculations. The FCIQMC and CCSD results, calculated within a DZVP basis set, are shown for both a finite supercell of 16 hydrogen atoms, and for the extrapolation to the infinite system size limit. Moreover, the CCSD-CBS energies, obtained using a finite simulation cell of 16 hydrogen atoms, are also illustrated. The shown DMC energies have also been extrapolated to the thermodynamic limit. Encouragingly, all three many-body wave function-based methods (FCIQMC, CCSD and DMC), which operate under very different assumptions and approximations, give the same equilibrium density for this phase, which is found to be close to $r_S=1.65$~a.u., with the finite-size correction for the DMC energies qualitatively reducing the equilibrium cell size. This agreement arises despite the fact that FCIQMC and CCSD are built from a Fock-space orbital representation, while DMC operates in the continuum, with the nodal surface being built on the Kohn-Sham DFT determinant. Yet, all three methods agree that the BCC atomic solid hydrogen with two atoms per cell has a minimum ground state energy at lattice parameter 3.307 Bohr. 

However, while the agreement with respect to the equilibrium geometry is good, it is clear that discrepancies between the approaches remain, with the $2\times2\times2$ supercell CCSD and FCIQMC results being in closer agreement with the thermodynamic limit DMC results than their extrapolated counterparts. This discrepancy is likely to be due to the extrapolation to the thermodynamic limit of CCSD, which was calculated only for the small SZV basis sets, which have previously been shown to be rather different to the larger DZVP basis sets. Further investigation into this effect and its mitigation is likely to be a continuing research direction.

\section{\label{con}Conclusion}
We report a numerical study of the EOS for BCC solid hydrogen within the density range of $1.3<r_S<2.4$~a.u., using FCIQMC, VMC, and DMC as many-body stochastic wave-function approaches, as well as the deterministic quantum chemical periodic CCSD method. We find that although sizable static correlation effects were captured at the FCIQMC level of theory for expanded cell sizes, at compressed geometries, however, these are less important and the results of CCSD and DMC agree rather well with those of FCIQMC. These complementary approaches can therefore be combined in order to mitigate their respective weaknesses, to provide a tractable route for the analysis of the electronic structure of extended systems. With careful consideration of basis set incompleteness and finite size errors, all methods demonstrated good agreement in the prediction of the equilibrium lattice parameter of 3.307 Bohr. However, discrepancies remain in the broader shape of the EOS, likely arising from the restricted basis sizes required for the thermodynamic limit extrapolation. Nevertheless, agreement between these methods, and demonstration of a viable composite scheme in these extended settings is likely a reliable approach to study further phases of these high-pressure hydrogenic materials.

\section{Acknowledgement}
S.A. gratefully acknowledges the generous allocation of computing time by the Paderborn Center for Parallel Computing (PC$^2$) on OCULUS and the FPGA-based supercomputer NOCTUA. G.H.B. has received funding from the Royal Society via a University Research Fellowship, as well as funding from the European Union's Horizon 2020 research and innovation program under grant agreement No. 759063. T.D.K. would like to the European Research Council (ERC) under the European Union’s Horizon 2020 Research and Innovation Programme (grant agreement no. 716142).

The data that support the findings of this study are available from the corresponding author upon reasonable request.


\end{document}